\documentclass[letterpaper]{jpconf}
\usepackage{iopams}
\usepackage[dvips]{graphicx}
\usepackage{pstricks}
\newcommand{\dif}{\mathrm{d}}
\newcommand{\gE}{\mathfrak{E}}
\newcommand{\gF}{\mathfrak{F}}
\renewcommand{\vec}{\bi}
\newcommand{\Atan}{\mbox{ArcTan\,}}
\newcommand{\Ti}{\mbox{Ti}_2\,}
\newcommand{\binom}[2]{ {#1 \choose #2} }
\begin{document}

\title{Weak Coupling Casimir Energies for Finite Plate Configurations}
\author{Jef Wagner, Kimball A. Milton, Prachi Parashar}

\address{
Oklahoma Center for High Energy Physics and
Homer L. Dodge Department of Physics and Astronomy,
University of Oklahoma, Norman, OK, 73019, USA}
\ead{wagner@nhn.ou.edu, milton@nhn.ou.edu, prachi@ou.edu}
\begin{abstract}
We derive and use an extremely simplified formula for the interaction 
Casimir energy for two separate bodies in the weak coupling regime 
for massless scalar fields. We derive closed form solutions for a general 
arrangement of two $\delta$-function plates finite in one direction 
and infinite in another. 
We examine the situation of two parallel plates finite in 
both transverse directions.
\end{abstract}
%\maketitle

\section{Introduction}
Recently, Emig et al.~\cite{EmigJaffe07, EGJK08} rederived the multiple 
scattering formalism and used it to calculate new results for Casimir
energies between disjoint bodies.
Similar techniques have been employed for many years,  perhaps starting 
with Renne in 1971 \cite{Renne71}, who rederived the Lifshitz 
formula \cite{Lifshitz55}. More recently many papers have used multiple
scattering techniques to examine the correction to the proximity force 
approximation for different situations
\cite{BMW06,WBM06,Bordag:2006kx,Bordag:2006vc,Bordag:2006zc,Milton:2007wz}.

Casimir calculations have conventionally been carried out assuming that 
parallel plates had infinite extent. Gies and Klingm\"uller have found
corrections to the infinite-size approximation for the cases of perfectly 
conducting boundary conditions \cite{Gies:2006xe}. We will here use the 
multiple scattering formalism to examine $\delta$-function plate geometries with finite lengths 
for scalar fields subject to very weakly coupled boundaries.

The weak coupling regime is interesting to study because the formula for the 
Casimir energy simplifies greatly, and in many cases is amenable to closed 
form solutions. This introduction gives a very quick derivation of the 
weak-coupling form of the Casimir energy for massless scalar fields. 
For a more complete derivation see Kenneth and Klich 
\cite{Kenneth:2006vr, Kenneth:2007jk}, 
or Milton and Wagner \cite{Milton:2007wz}.

The vacuum expectation value for the action of a scalar field can be given 
by the standard formula
\begin{equation}\label{VacAction}
W=-\frac{\rmi}{2}\Tr \ln \frac{\mathcal{G}}{\mathcal{G}_0},
\end{equation}
where $\mathcal{G}$ is the Green's function that satisfies the wave equation 
of the scalar field, including any interactions with any objects or 
potentials. $\mathcal{G}_0$ is the free Green's function that satisfies the 
wave equation with the same boundary conditions at infinity as the full 
Green's function but without any interaction with background potentials. 
For a time-independent system we can use the condition that 
$W=-\int \dif t E$ to identify the energy as
\begin{equation}
E=\frac{\rmi}{2}\int \frac{\dif \omega}{2 \pi}
\Tr \ln \frac{G}{G_0},
\end{equation}
where $G$ is the Fourier-transformed Green's function given by
\begin{equation}
\mathcal{G}(x,x')=\int \frac{\dif \omega}{2 \pi}
\rme^{-\rmi \omega (t-t')}G(\vec{x},\vec{x}'),
\end{equation}
and the trace if over spatial coordinates.
Given two nonoverlapping potentials such that $V(x)=V_1(x)+V_2(x)$, 
we can define an interaction energy as the energy of the full system 
less the energy of each potential acting alone, 
$E_{\rm Int} = E(V_1+V_2) - E(V_1) - E(V_2)$. This simplified expression 
can be written as
\begin{equation}\label{EnMScat}
E_{\rm Int}=-\frac{\rmi}{2}\int \frac{\dif \omega}{2 \pi}
\Tr \ln \left(1 - G_0 T_1 G_0 T_2 \right),
\end{equation}
where $T_i$ is the scattering matrix for the $i^{th}$ potential defined as 
$T_i=V_i(1-G_0 V_i)^{-1}$.

By formally expanding out the logarithm 
in \eref{EnMScat}, we can think about 
the formula as describing successively more scattering events between the 
two objects. The weak coupling expansion keeps only the first term of the 
expansion of the logarithm, essentially describing only a single scattering 
between the objects. Additionally, in the weak-coupling regime the scattering 
matrix can be approximated simply by the potential, $T \approx V$. This 
results in a very simplified weak-coupling single-scattering approximation to 
the energy ($\omega\to\rmi\zeta$),
\begin{equation}\label{EnWScat}
E_{\rm Int}=-\frac{1}{4 \pi}\int \dif \zeta
\Tr G_0 V_1 G_0 V_2.
\end{equation}

\section{2+1 Spatial Dimensions}
If the potentials are independent of the $z$ direction then we can further 
simplify the interaction energy. By dividing out the infinite length in the 
$z$ direction we obtain an energy per unit length, which we will represent 
in this paper by the fraktur symbol $\gE$. By Fourier transforming in the $z$ 
direction we get
\begin{equation}
\gE=
-\frac{1}{2}\int \frac{\dif \zeta}{2\pi}
\int \frac{\dif k_z}{2\pi} 
\Tr g_0 V_1 g_0 V_2,
\end{equation}
where $g_0$ is given by
\begin{equation}
G_0(\vec{r}-\vec{r}';\zeta)=\int \frac{\dif k_z}{2 \pi}
\rme^{\rmi k_z (z-z')}g_0(\vec{r}_\perp-\vec{r}'_\perp;\kappa),
\end{equation}
and $\kappa$ is defined by $\kappa^2=\zeta^2+k_z^2$. The two dimensional 
Green's function is explicitly
\begin{equation}
g_0(\vec{r}_\perp-\vec{r}'_\perp;\kappa)=\frac{1}{2\pi}
K_0(\kappa |\vec{r}_\perp-\vec{r}'_\perp|),
\end{equation}
yielding a weak-coupling form for the energy per unit length of
\begin{equation}\label{enWScat2d_k}
\gE=-\frac{1}{32 \pi^4}\int \dif \zeta \dif k_z
\int \dif^2 r \int \dif^2 r' 
K_0^2(\kappa |\vec{r}_\perp-\vec{r}'_\perp|)
V_1(\vec{r}_\perp;\zeta)V_2(\vec{r}'_\perp;\zeta).
\end{equation}
In the case that the potentials are independent of the imaginary frequency 
$\zeta$ this simplifies even further to
\begin{equation}\label{enWScat2d}
\gE=-\frac{1}{32 \pi^3}\int \dif^2 r \int \dif^2 r'
\frac{V_1(\vec{r}_\perp)V_2(\vec{r}'_\perp)}
{|\vec{r}_\perp-\vec{r}'_\perp|^2}.
\end{equation}

\begin{figure}[h]
\centering
\includegraphics{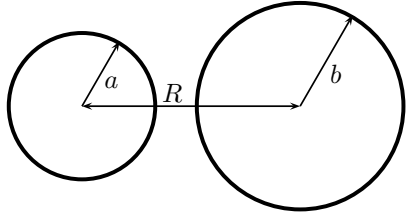}
%\begin{pspicture}(6,3)
%  \psgrid
%  \psset{linewidth=1.5pt}
%  \pscircle(1.5,1.5){1}
%  \psset{linewidth=.65pt}
%  \psline{->}(1.5,1.5)(2,2.366025404)
%  \rput[tl]{0}(1.8,1.9){$a$}
%  \psset{linewidth=1.5pt}
%  \pscircle(4.4,1.5){1.4}
%  \psset{linewidth=.65pt}
%  \psline{->}(4.4,1.5)(5.1,2.7124355656)
%  \rput[tl]{0}(4.8,2.05){$b$}
%  \psline{<->}(1.5,1.5)(4.4,1.5)
%  \rput[b]{0}(2.7,1.55){$R$}
%\end{pspicture}
\caption{\label{Cylinders}Two cylinders of radii $a$ and $b$, their 
centers separated by a distance $R$.}
\end{figure}

To demonstrate the simplicity of this formula  we will rederive the 
Casimir energy for two cylinders \cite{Milton:2007wz}. Assume that the two 
cylinders have radii $a$ and $b$, and their centers are separated by a 
distance $R$ as shown in \fref{Cylinders}.
This situation can be represented by potentials $V_1=\lambda_1 \delta(r-a)$ 
and $V_2=\lambda_2 \delta(r'-b)$ where $r$ and $r'$ are radial coordinates 
in cylindrical polar coordinate systems centered on the respective cylinders. 
Using \eref{enWScat2d},  these potentials yield an energy per unit length
\begin{equation}
\gE=-\frac{\lambda_1 \lambda_2 a b}{32 \pi^3}
\int_0^{2\pi}\dif \theta \int_0^{2\pi} \dif \theta'
\frac{1}{R^2+a^2+b^2-2aR \cos \theta + 
2bR \cos \theta'-2ab \cos (\theta-\theta')}.
\end{equation}
With a simple change in angular coordinates to $u=\theta-\theta'$ and 
$v=\frac{\theta+\theta'}{2}$ this expression can be integrated to yield 
the exact closed form \cite{Milton:2007wz}
\begin{equation}
\gE=-\frac{\lambda_1 \lambda_2 a b}{8\pi}
\frac{1}{\sqrt{\left(R^2-(a-b)^2\right)
\left(R^2-(a+b)^2\right)}}.
\end{equation}

\subsection{General Configuration of Plates}
\begin{figure}[h]
\centering
\includegraphics{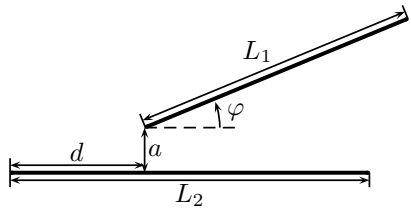}
%\begin{pspicture}(6,3)
%  \psgrid
%  \psset{linewidth=1.5pt}
%  \psline{cc-cc}(.2,.4)(5,.4)
%  \psline{cc-cc}(2,1)(5.51074, 2.4542)
%  \psset{linewidth=.65pt}
%  \psline{|<->|}(.2,.3)(5,.3)
%  \rput[t]{0}(2.6,.25){$L_2$}
%  \psline{|<->|}(1.96173, 1.09239)(5.47247, 2.54658)
%  \rput[br]{0}(3.69797, 1.86568){$L_1$}
%  \psline{|<->}(.2,.5)(2,.5)
%  \rput[b]{0}(1.1,.55){$d$}
%  \psline{<->}(2,.4)(2,1)
%  \rput[l]{0}(2.05,.7){$a$}
%  \psline[linestyle=dashed]{-}(2,1)(3.2,1)
%  \psarc[arcsepB=2.5pt]{->}(2,1){1}{0}{22.5}
%  \rput[bl]{0}(3.1,1.1){$\varphi$}
%\end{pspicture}
\caption{\label{2Dgenfig}Two finite plates of length $L_1$ and $L_2$ in a 
general configuration. In a coordinate system centered at the edge of the top 
plate $\varphi$ is the relative angle between the plates, $a$ is the 
perpendicular distance the lower plate is shifted down, and $d$ is the lateral
distance the edge of the lower plate is shifted to the left.}
\end{figure}

Two finite plates of a general configuration as shown in \fref{2Dgenfig} 
can be represented by the potentials in cylindrical coordinates,
with origin at the left edge of plate 1,
\numparts
\begin{eqnarray}
V_1=\lambda_1 \delta(\theta-\varphi)\frac{\Theta(L_1-r)}{r},\\
V_2=\lambda_2 \delta(y+a)\Theta(x+d)\Theta(L_2-d-x).
\end{eqnarray}
\endnumparts
Here $\Theta(x)$ is the step function,
\begin{equation}
\Theta(x)=\left\{\begin{array}{cc}
1,&x>0,\\
0,&x<0.\end{array}\right.
\end{equation}
Using these potentials with \eref{enWScat2d}, we get the following 
expression for the energy per unit length,
\begin{equation}
\gE=-\frac{\lambda_1 \lambda_2}{32 \pi^3}
\int_0^{L_1} \dif r \int_{-d}^{L_2-d} \dif x \frac{1}
{(x-r \cos \varphi)^2+(a+r \sin \varphi)^2}.
\end{equation}
This integral can be done exactly, yielding a closed form for the general 
configuration,
\begin{eqnarray}\label{2Dgen}
\gE=-\frac{\lambda_1\lambda_2}{32 \pi^3 \sin \varphi}
\left[
\Ti\left( \frac{L_2-d}{a}, \cot \varphi \right)
-\Ti\left( \frac{L_2-d-L_1 \cos \varphi}
{a+L_1 \sin \varphi}, \cot \varphi \right)
\right.\nonumber\\
\left.
-\Ti\left( \frac{-d}{a}, \cot \varphi \right)
+\Ti\left( \frac{-d-L_1 \cos \varphi}{a+L_1 \sin \varphi},
\cot \varphi \right)
\right],
\end{eqnarray}
where $\Ti$ is the generalized inverse tangent integral\footnote{The 
generalized inverse tangent integral is related to the dilogarithm function, 
and much information about it can be found in \cite{Lewin81}.}
defined by
\begin{equation}
\Ti(x,a)=\int_0^x \dif y \frac{\Atan y}{y+a}.
\end{equation}

\subsection{Torque}
\begin{figure}[h]
\centering
\includegraphics{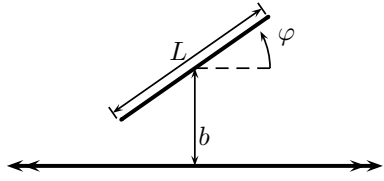}
%\begin{pspicture}(6,3)
%  \psset{linewidth=1.5pt}
%  \psline{<<->>}(.5,.4)(5.5,.4)
%  \psline{cc-cc}(2,1)(4,2.4)
%  \psset{linewidth=.65pt}
%  \psline{|<->|}(1.9,1.1)(3.9,2.5)
%  \rput[br]{0}(2.9,1.8){$L$}
%  \psline{<->}(3,.4)(3,1.7)
%  \rput[l]{0}(3.05,.8){$b$}
%  \psline[linestyle=dashed]{-}(3,1.7)(4,1.7)
%  \psarc[arcsepB=1.5pt]{->}(3,1.7){1}{0}{31}
%  \rput[bl]{0}(4.1,2){$\varphi$}
%\end{pspicture}
\caption{\label{torquefig}A finite plate of length $L$, above an infinite 
plate, The center of the finite plate is a perpendicular distance $b$ above 
the infinite plate, and the finite plate makes an angle $\varphi$ with 
respect to the infinite plate.}
\end{figure}
It would be desirable to examine the rotational 
stability of parallel plates. This 
can be done simply by looking at the sign of the torque for small angular 
displacements, while leaving the center of mass stationary. Although one can 
get a general form for the torque from \eref{2Dgen}, it is much simpler 
to restructure the problem to isolate the torque as shown in 
\fref{torquefig}. This is the same as the ``Casimir pendulum'' problem studied by Scardicchio et al., who used the optical approximation 
\cite{Scardicchio:2004fy}. 

Given a situation of a tilted plate of length $L$ over an infinite plate
($L_1\to L$, $L_2\to\infty$, $d\to-\infty$ in \fref{2Dgenfig}), 
we can very easily 
isolate the torque that the finite plate experiences. The energy per unit 
length is given by
\begin{equation}
\gE=-\frac{\lambda_1 \lambda_2}{32 \pi^2 \sin \varphi}
\ln \left( \frac{b+\frac{L}{2}\sin \varphi}
{b-\frac{L}{2}\sin \varphi}\right).
\end{equation}
The torque per unit length $\mathfrak{T}$ found by taking the negative  
derivative of the energy with respect to the tilt angle. This gives an 
expression for the torque on the plates as
\begin{equation}\label{torque}
\mathfrak{T}=-\frac{\lambda_1 \lambda_2}{32 \pi^2}
\frac{\cos \varphi}{\sin \varphi}
\left(\frac{1}{\sin \varphi}
\ln \left( \frac{b+\frac{L}{2}\sin \varphi}
{b-\frac{L}{2}\sin \varphi}\right)
-\frac{L b}{b^2-\frac{L^2}{4}\sin^2 \varphi}
\right).
\end{equation}

From the expression for the torque we can see some clear qualitative features. 
The torque has a zero value at $\varphi=0$, and a quick evaluation shows 
that the first derivative is positive, of value
\begin{equation}
\left.\frac{\partial\mathfrak{T}}
{\partial\varphi}\right|_{\varphi=0}=
\frac{\lambda_1 \lambda_2}{192 \pi^2}
\frac{L^3}{b^3},
\end{equation}
signifying an unstable equilibrium. For values of $L$ such that $L>2b$, the 
torque diverges as $\sin \varphi$ approaches $2b/L$. This is simply the result 
of the fact that the plates would touch in this situation. If $L<2b$ the torque
has another zero at $\varphi=\frac{\pi}{2}$, with a first derivative of
\begin{equation}
\left.\frac{\partial\mathfrak{T}}
{\partial\varphi}\right|_{\varphi=\frac{\pi}{2}}=
-\frac{\lambda_1 \lambda_2}{32 \pi^2}
\left(\frac{L b}{b^2-\frac{L^2}{4}}
-\ln \left( \frac{b+\frac{L}{2}}
{b-\frac{L}{2}}\right)
\right),
\end{equation}
which is negative for all $L<2b$, meaning a stable equilibrium. Therefore a 
finite flat plate suspended above another infinite plate will tend to orient 
itself perpendicular to the plate if left to rotate about its center of mass.
Physically, the reason for this is clear: because the magnitude of
the Casimir force
decreases with distance, to minimize the energy for a fixed center of mass,
the smaller plate wants to rotate so as to place the shortest side closest
to the infinite plate. 
(For the situation when the thickness of the plate is finite,
see \cite{Milton:2008vr}.)

\subsection{\label{PPsec}Parallel Plates}
\begin{figure}[h]
\centering
\includegraphics{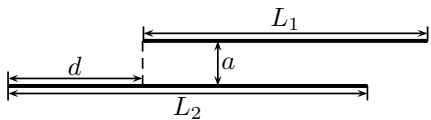}
%\begin{pspicture}(6,1.5)
%  \psgrid
%  \psset{linewidth=1.5pt}
%  \psline{cc-cc}(.2,.4)(5,.4)
%  \psline{cc-cc}(2,1)(5.8,1)
%  \psset{linewidth=.65pt}
%  \psline{|<->|}(.2,.3)(5,.3)
%  \rput[t]{0}(2.6,.25){$L_2$}
%  \psline{|<->|}(2,1.1)(5.8,1.1)
%  \rput[b]{0}(3.9,1.15){$L_1$}
%  \psline{|<->}(.2,.5)(2,.5)
%  \rput[b]{0}(1.1,.55){$d$}
%  \psline[linestyle=dashed]{-}(2,.4)(2,1)
%  \psline{<->}(3,.4)(3,1)
%  \rput[l]{0}(3.05,.7){$a$}
%\end{pspicture}
\caption{\label{PPfig}Parallel plates}
\end{figure}

Parallel plates are perhaps the most interesting special case. We can compare 
the exact expressions for energy and force to those for infinite parallel 
plates, getting corrections for finite size. In addition the parallel plates 
case, due to its simplicity, lends itself well to studying both normal and 
lateral forces.

Consider the same setup as in the general case shown in \fref{2Dgenfig}, 
simply letting $\varphi$ go to zero, as shown in figure \fref{PPfig}. 
The energy per unit length can be derived 
directly from \eref{2Dgen}  by using the identity
\begin{equation}
\lim_{a \to \infty} a \Ti(x,a)=\int \dif x \Atan x,
\end{equation}
yielding an integral form for the energy per unit length
\begin{equation}\label{enPP}
\gE=-\frac{\lambda_1 \lambda_2}{32 \pi^3}
\left[
\int\limits_{\frac{L_2-d-L_1}{a}}^{\frac{L_2-d}{a}} +
\int\limits_{\frac{d}{a}}^{\frac{d+L_1}{a}} 
\right] \dif x 
\Atan x .
\end{equation}
Although an indefinite integral for the arctangent exists, this form is 
perhaps more illuminating because all the physical quantities are in the 
limits.
The forces, which are given as derivatives of the energy, are all given in 
terms of arctangents.

Equation \eref{enPP} yields closed forms for the normal force between the 
plates and the lateral force experienced by the plates by taking the negative 
derivative with respect to $a$ or $d$, respectively. The general form of the 
normal force, defined as $\gF_a=-\partial\gE/\partial a$, is
\begin{eqnarray}\label{2Dforceat}
\gF_a=-\frac{\lambda_1 \lambda_2}{32 \pi^3 a^2}
\left[
(L_2-d)\Atan\left(\frac{L_2-d}{a}\right)
-(L_2-d-L_1)\Atan\left(\frac{L_2-d-L_1}{a}\right)
\right.\nonumber\\
\left.
-d\Atan\left(\frac{d}{a}\right)
+(d+L_1)\Atan\left(\frac{d+L_1}{a}\right)
\right].
\end{eqnarray}
In the limiting case of the plates getting very close together we expect to 
recover the result for the pressure for infinite parallel plates times the 
area exposed. By mathematically taking $a \to 0$, we use the large argument 
expansion of the inverse tangent, 
\begin{equation}\label{atan}
\Atan(x)=
\frac{\pi}{2} - \frac{1}{x} + \frac{1}{3}\frac{1}{x^3}+\cdots, 
\qquad {\rm for}\; x\to \infty,
\end{equation}
to recover the expected result plus corrections to that result. Because the 
limiting form of the arctangent depends on the sign of the argument, the 
single general equation can give several different answers depending on the 
size and position of the plates. For the situation shown in \fref{PPfig} 
the limiting form is
\begin{equation}
\gF_a=-\frac{\lambda_1 \lambda_2}{32 \pi^2 a^2}
\left( (L_2-d)+\mathcal{O}(a^3) \right) ,
\end{equation}
and the first correction is zero. However, if the plates are the same size 
and aligned the limiting form of the force with the first correction is
\begin{equation}
\gF_a=-\frac{\lambda_1 \lambda_2}{32 \pi^2 a^2}
\left(L-\frac{1}{\pi}2a+\mathcal{O}(a^3)\right).
\end{equation}
If we let one end of both plates extend off into infinity then we can get 
the edge correction for two aligned plates. This correction is
\begin{equation}\label{edgecor} 
\frac{\gF_a/\gF_0-1}{2a}=\frac{1}{\pi}.
\end{equation}
The general form of the lateral force, similarly defined as 
$\gF_d=-\partial\gE/\partial d$, is
\begin{eqnarray}\label{2Dforcelat}
\gF_d=-\frac{\lambda_1 \lambda_2}{32 \pi^3 a}
\Bigg[
\Atan\left(\frac{L_2-d-L_1}{a}\right)
-\Atan\left(\frac{L_2-d}{a}\right)\nonumber\\
-\Atan\left(\frac{d}{a}\right)
+\Atan\left(\frac{d+L_1}{a}\right)
\Bigg].
\end{eqnarray}
From the exact expression for the lateral force, we find there is only one 
equilibrium position, occurring at $d=\frac{L_1-L_2}{2}$, where
the derivative of the force is  negative:
\begin{equation}
\left.\frac{\partial \gF_d}{\partial d}\right|_{d=\frac{L_1-L_2}{a}}
=-\frac{\lambda_1 \lambda_2}{16 \pi^3}
\frac{L_1 L_2}{
\left(a^2+\left(
  \frac{L_1+L_2}{2}
\right)^2 \right)
\left(a^2+\left(
  \frac{L_1-L_2}{2}
\right)^2 \right) },
\end{equation}
signifying a stable equilibrium. The position and qualitative behavior is as 
expected, the plate have an stable equilibrium when they are symmetrically 
aligned.

We are also interested in how the lateral force behaves if the plates are very close together. To study that we simply take the limit as $a \to 0$. 
Assuming without loss of generality that $L_2>L_1$, 
to lowest order the force is
\begin{equation}
\gF_d=\left\{\begin{array}{ll}
+\frac{\lambda_1 \lambda_2}{16 \pi^2 a} &
\qquad {\rm for}\; d>0 \; {\rm and}\; d>L_2-L_1\\
0 & \qquad {\rm for} \; d>0 \; {\rm and}\; 0<d<L_2-L_1\\
-\frac{\lambda_1 \lambda_2}{16 \pi^2 a} &
\qquad {\rm for} \; d<0
\end{array}\right.
\end{equation}
This is what we would expect if we approximated the 
energy simply as the energy per area between the two infinite plates 
times the area exposed between the two plates, and took the derivative 
of this very simple approximation as the force.

\section{Three spatial dimensions}
Until now we have worked in 2+1 dimensions, meaning that the potentials had 
infinite length in one direction. So the finite plates considered in the 
previous section were more like two infinite ribbons. In this section we 
will work with plates of finite area.

We start by working out the $TGTG$ formula \eref{EnWScat} in three dimensions. 
In three dimensions the form of the Green's function is even easier to work 
with than in two dimensions,
\begin{equation}
G_0(\vec{r},\vec{r'})=\frac{1}{4\pi}\frac{e^{-|\zeta||\vec{r}-\vec{r'}|}}
{|\vec{r}-\vec{r'}|}.
\end{equation}
This gives a form of the energy
\begin{equation}
E=-\frac{1}{64 \pi^3}
\int_{-\infty}^\infty \dif \zeta \int \dif^3 r \int \dif^3 r'
\frac{e^{-2|\zeta||\vec{r}-\vec{r'}|}
V_1(\vec{r};\zeta)V_2(\vec{r'};\zeta)}
{|\vec{r}-\vec{r'}|^2},
\end{equation}
and if the potentials are independent of $\zeta$ then the erpression simplifies
further to
\begin{equation}
E=-\frac{1}{64 \pi^3}
\int \dif^3 r \int \dif^3 r'
\frac{V_1(\vec{r})V_2(\vec{r'})}
{|\vec{r}-\vec{r'}|^3}.
\end{equation}

If we restrict ourselves to parallel plate $\delta$-function potentials of any 
general shape the energy can written in an even simpler form,
\begin{equation}
E=-\frac{\lambda_1 \lambda_2}{64 \pi^3}
\int_{A_1} \dif^2 r_\perp \int_{A_2} \dif^2 r'_\perp
\left[ a^2+(\vec{r}_\perp-\vec{r'}_\perp )^2 \right] ^{-3/2},
\end{equation}
where $A_1$ and $A_2$ are the areas of the two plates,
and $a$, again, is the separation between the plates. If we let one of the 
areas, let it be $A_2$, tend to infinity, then the energy for a single finite 
plate above an infinite plate is
\begin{equation}
E=-\frac{\lambda_1 \lambda_2}{32 \pi^2 a}A_1,
\end{equation}
exactly the energy per area from the Lifshitz formula times the area of the 
finite plate. For weak coupling plates, if one of the plates is infinite, and 
the other finite, then there is no correction to the Lifshitz formula. This 
is not unexpected, and is a result of the fact that the weak coupling 
approximation is the same as pairwise summation.

\subsection{Rectangular Parallel Plates}
\begin{figure}[h]
\centering
\includegraphics{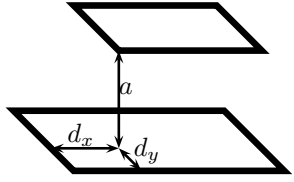}
%\begin{pspicture}(4,3)
%  \psgrid
%  \psset{linewidth=2.5pt}
%  \pspolygon(1,0.4)(3.8,0.4)(3,1.2)(0.2,1.2)
%  \pspolygon(1.6,2)(3.5,2)(2.9,2.6)(1,2.6)
%  \psset{linewidth=1pt}
%  \psline{<->}(1.6,0.7)(1.6,2)
%  \rput[l]{0}(1.6,1.5){$a$}
%  \psline{<->}(1.9,0.4)(1.6,0.7)
%  \rput[bl]{0}(1.8,0.5){$d_y$}
%  \psline{<->}(0.7,0.7)(1.6,0.7)
%  \rput[b]{0}(1.1,0.73){$d_x$}
%\end{pspicture}
\caption{\label{2pp} Two finite parallel rectangular plates. 
The sizes of the plates are 
$L_{1x} \times L_{1y}$ for the top plate, and $L_{2x} \times L_{2y}$ for the 
bottom plate. The lower left corner of the upper plate is displaced a 
distance $d_x$ in the $x$ direction and $d_y$ in the $y$.}
\end{figure}

For two rectangular parallel plates, as shown in \fref{2pp}, 
the interaction energy is given by the integral
\begin{equation}
E=-\frac{\lambda_1 \lambda_2}{64 \pi^3}
\int\limits_{d_x}^{L_{1x}+d_x}\dif x \int\limits_0^{L_{2x}} \dif x'
\int\limits_{d_y}^{L_{1y}+d_y}\dif y \int\limits_0^{L_{2y}} \dif y'
\left[a^2+(x-x')^2+(y-y')^2\right]^{-3/2}.
\end{equation}
This expression can be partially integrated and rewritten as
\begin{equation}
  E=\frac{-\lambda_1 \lambda_2 a}{64 \pi^3}
  \left[
    \int\limits_{\frac{ L_{1x}+d_x-L_{2x}}{a}}^{\frac{L_{1x}+d_x}{a}} +
    \int\limits_{\frac{d_x}{a}}^{\frac{d_x-L_{2x}}{a}} 
    \right] \dif x
  \left[
    \int\limits_{\frac{L_{1y}+d_y-L_{2y}}{a}}^{\frac{L_{1y}+d_y}{a}} +
    \int\limits_{\frac{d_y}{a}}^{\frac{d_y-L_{2y}}{a}} 
    \right] \dif y
  \Atan\left( \frac{xy}{\sqrt{1+x^2+y^y}} \right).
\end{equation}
The two-dimensional indefinite integral in the equation is given by 
\begin{eqnarray}
\int \dif x \int \dif y 
\Atan \left( \frac{xy}{\sqrt{1+x^2+y^2}} \right) =
x y \Atan \left( \frac{xy}{\sqrt{1+x^2+y^2}} \right) \nonumber\\
+x \ln\left(x+\sqrt{1+x^2+y^2}\right)
+y \ln\left(y+\sqrt{1+x^2+y^2}\right) \nonumber \\
-\frac{1}{2}x\ln(1+y^2)-\frac{1}{2}y\ln(1+x^2)-\sqrt{1+x^2+y^2}.
\end{eqnarray}
The final closed-form expression for the energy of the two rectangular 
parallel plates is somewhat messy, consisting of the above indefinite integral 
evaluated at 16 different combinations of variables.

The normal and lateral forces can again be given by the derivatives of the 
energy with respect to the separation $a$ or to the displacement 
(this time either $d_x$ or $d_y$).

The lateral force from the plates has a stable equilibrium when the centers of
the two plates are aligned. However, first derivatives of the force can be 
different for displacements from the equilibrium position in the $x$ and $y$ 
directions depending on the geometry.

Perhaps the most interesting property of this system as in section \ref{PPsec}
to examine what 
happens to the attractive force between the plates as the plates get very 
close together. For very small separations we should get an expression for the 
force as a power series in $a$ where the first term is the pressure given by 
the Lifshitz formula times the area between the plates,
\begin{equation}
F_a=-\frac{\lambda_1 \lambda_2}{32 \pi^2 a^2}A(1+c_1 a + c_2 a^2 \cdots ).
\end{equation}
Using the large argument expansion for the arctangent \eref{atan}, 
it is possible to get 
such an expression for the two plate arrangement, although the expressions 
for the area and the correction terms depend on the layout of the plates. 
For a situation in which the upper plate is completely above the 
lower plate, with none of the edges aligned, the area is given as 
$A=L_{1x} L_{1y}$ and the first correction term is $c_1=0$. For a situation 
where both plates are the same size, and they are exactly aligned ($d_x=d_y=0$)
then the area is $A=L_x L_y=L_{1x} L_{1y}=L_{2x} L_{2y}$ and 
\begin{equation}
c_1=-\frac{1}{\pi}\frac{2(L_x+L_y)}{L_x L_y}.
\end{equation}

\subsection{Parallel Disks}
\begin{figure}
\centering
\includegraphics{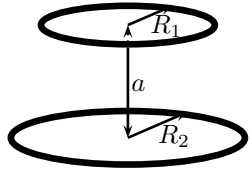}
%\begin{pspicture}(4,3)
%  \psgrid
%  \psset{linewidth=2.5pt}
%  \psellipse(2,0.8)(1.6,0.4)
%  \psellipse(2,2.3)(1.2,0.3)
%  \psset{linewidth=1pt}
%  \psline{->}(2,0.8)(2.8, 1.14641)
%  \rput[tl]{0}(2.4,0.973205){$R_2$}
%  \psline{->}(2,2.3)(2.6, 2.55981)
%  \rput[tl]{0}(2.3,2.4299){$R_1$}
%  \psline{<-}(2,0.8)(2,2)
%  \psline[linestyle=dotted,dotsep=1.5pt]{->}(2,2)(2,2.3)
%  \rput[l]{0}(2.05,1.5){$a$}
%\end{pspicture}
\caption{Two coaxial disks, separated by a distance $a$. The radii for the two 
disks are $R_1$ and $R_2$.}
\end{figure}

Instead of asking how two parallel rectangular plates attract, we could just 
have easily asked how two disks attract. The interaction Casimir energy for 
two coaxial disks separated by a distance $a$ is given by 
\begin{equation}
E=-\frac{\lambda_1\lambda_2}{32 \pi^2}
\int_0^{R_1} \dif r \int_0^{R_2} \dif r' \int_0^{2\pi} \dif \theta
\frac{r r'}{\left[ a^2 +r^2 +r'^2 -2 r r' \cos\theta \right]^{3/2}}.
\end{equation}

This expression can be integrated term by term in a series expansion in powers 
of $r$ and $r'$. The energy can then be expressed as
\begin{equation}
E=-\frac{\lambda_1\lambda_2}{32 \pi}\frac{R_1^2 R_2^2}{a^3} 
\sum_{m=0}^\infty \sum_{n=0}^m A_{m,n}
\left(\frac{R_1}{a}\right)^{2(m-n)}
\left(\frac{R_2}{a}\right)^{2n},
\end{equation}
where
\begin{equation}
A_{m,n}= \frac{1}{2}\left(\frac{-1}{4}\right)^m
\binom{2(m+1)}{m+1}
\left[ \binom{m}{n}^2
-\binom{m}{n+1} \binom{m}{n-1} \right].
\end{equation}

The power series is convergent, so we can simply take the derivative of each 
term to get the force between the plates. By using the asymptotics of the power
series in the limit as $a \to 0$ we recover the expected result and get 
corrections to the infinite plate result. For two different sized disks, if 
$R_1<R_2$ then $A=\pi R_1^2$ and $c_1=0$. For two equal sized disks where 
$R_1=R_2=R$, $A=\pi R^2$ and
\begin{equation}
c_1=-\frac{1}{\pi}\frac{2\pi R}{\pi R^2}.
\end{equation}

\section{Conclusions}
The weak-coupling regime greatly simplifies Casimir calculations, and even more
so the interaction energy between two bodies. The simplicity allows us to
obtain closed-form solutions to the energy and force between some nontrivial 
geometries. These closed-form solutions can help us in understanding the 
mechanics of these systems, such as the torque on two finite plates studied 
here. Also these solutions can help us in understanding the limits of certain 
approximations, such as the proximity force approximation or the correction to 
the attractive force between plates of finite size.

For the three cases of parallel plates studied here, the 2+1 D parallel plates,
the rectangular parallel plates, and the parallel co-axial disks, we get very 
similar results in the limit as the plates get very close together.
If the edges of the plates do not align then the first correction term 
$c_1=0$ and the area is the area of overlap of the plates. 
This fact can be reconciled with the fact that for a finite 
plate over an infinite plate the exact result for the attractive force is 
simply the pressure from the Lifshitz formula times the area of the plate. We 
would expect this to be a good approximation if one plate were much larger than
the other, which corresponds to the first correction being zero. If one plate 
is even slightly larger, in the limit as the separation goes to zero the 
difference in the size of the plates is still large in comparison to the 
separation.

In the case of the edges of the two plates aligning exactly
(that is, plates of the same size and shape directly above one another) 
then the area is simply the area of the plates and
the first correction term takes the form
\begin{equation}
c_1=-\frac{1}{\pi}\frac{\rm Perimeter}{\rm Area}.
\end{equation}
This is a general property of this system, and can be visualized by 
realizing that in the limit as $a$ goes to zero locally along any edge the 
system will appear to be two semi-infinite plates with their edges aligned. 
%for which the edge correction is given by \eref{edgecor}. 
Therefore we might expect the correction to be a proportional to the perimeter 
of the plates, with the constant of proportionality given by \eref{edgecor}.

In addition, these closed-form results will act as simple test cases for 
numerical studies.
\ack
%\section{Acknowledgements}
We thank K.V. Shajesh and S.A. Fulling for helpful remarks.  This
work is supported in part by the US National Science Foundation
and the US Department of Energy.

\section*{References}

\bibliography{finite3}

\end{document}